\documentclass{birkmult}

\usepackage{amssymb}

\newtheorem{theorem}{Theorem}[section]

\newtheorem{Pro}[theorem]{Proposition}
\newtheorem{Lem}[theorem]{Lemma}
\newtheorem{Cor}[theorem]{Corollary}
\newtheorem{Deff}[theorem]{Definition}

\newcommand{\fa}{\forall}
\newcommand{\Ga}{\Gamma}
\newcommand{\Gas}{\Gamma^\star}
\newcommand{\Gao}{\Gamma^\omega}

\newcommand{\Si}{\Sigma}
\newcommand{\Sis}{\Sigma^\star}
\newcommand{\Sio}{\Sigma^\omega}
\newcommand{\ra}{\rightarrow}
\newcommand{\hs}{\hspace{12mm}

\noi}
\newcommand{\lra}{\leftrightarrow}

\newcommand{\ite}{\item}

\newcommand{\ol}{ $\omega$-language}
\newcommand{\orl}{ $\omega$-regular language}
\newcommand{\om}{\omega}
\newcommand{\nl}{\newline}
\newcommand{\noi}{\noindent}

\newcommand{\proo}{\noi {\bf Proof.} }
\newcommand {\ep}{\hfill $\square$}

\def\ufootnote#1{\let\savedthfn\thefootnote\let\thefootnote\relax
\footnote{#1}\let\thefootnote\savedthfn\addtocounter{footnote}{-1}}

\begin{document}

\title{{\bf Wadge Degrees of Infinitary Rational Relations$^\star$  }
 }

\author{Olivier Finkel}
\address{{\it Equipe Mod\`{e}les de Calcul et Complexit\'e}  
 \\ {\it Laboratoire de l'Informatique du Parall\'elisme } 
\\ UMR 5668 - CNRS - ENS Lyon - UCB Lyon - INRIA
 \\  CNRS et Ecole Normale Sup\'erieure de Lyon
 \\ 46, All\'ee d'Italie 69364 Lyon Cedex 07, France. }
\address{ and \\ {\it Institut des Syst\`{e}mes complexes} \\ 5, Rue du Vercors \\
69007 Lyon, France.}

\email{Olivier.Finkel@ens-lyon.fr}

\date{}

\ufootnote{$\star$ LIP Research Report RR 2008-14}

\ufootnote{$\star$ This paper is an extended version of a conference paper  which appeared in the Proceedings 
of  the 23rd International Symposium on Theoretical Aspects of Computer Science, 
STACS 2006, \cite{Fin06b}.}

\begin{abstract}
\noi 
We show that, from the topological point of view,  $2$-tape B\"uchi automata  have the same accepting power as 
Turing machines equipped with a  B\"uchi acceptance condition. 
The Borel and the Wadge hierarchies  of the class ${\bf RAT}_\om$ of infinitary rational relations accepted by $2$-tape B\"uchi automata are 
 equal to the Borel and the Wadge hierarchies  of 
$\om$-languages accepted  by real-time B\"uchi $1$-counter automata or by B\"uchi Turing machines. 
In particular,   for every non null recursive ordinal $\alpha$,  
there exist some  
${\bf \Si}^0_\alpha$-complete and some ${\bf \Pi}^0_\alpha$-complete infinitary rational relations.   And the supremum of the set of Borel ranks of 
infinitary rational relations is an ordinal $\gamma_2^1$ which is strictly greater than the first non recursive ordinal   
$\om_1^{\mathrm{CK}}$. 
This  very surprising result gives answers to 
questions of Simonnet \cite{Simonnet92} and of Lescow and Thomas \cite{Thomas88,LescowThomas}. 
\end{abstract}


\maketitle

{\bf  \hfill   Dedicated to Bill Wadge on the occasion of his 60 th birthday}

\section{Introduction}

In the sixties, automata accepting  infinite words were  firstly 
considered by B\"uchi in order to study 
decidability of the monadic second order theory S1S
of one successor over the integers \cite{Buchi62}. 
Then the so called \orl s have been 
intensively studied and have found many applications for 
specification and verification of non terminating systems, see \cite{Thomas90,Staiger97,PerrinPin} for many results and references. 
 On the other hand, rational relations on finite words were  also studied in the sixties,  and played 
a fundamental role in the study of families of context free languages \cite{Berstel79}. 
Investigations on their extension to rational  relations on infinite words were carried out 
or mentioned in the books \cite{BarzdinTrakhtenbrot,LindnerStaiger}. Gire  and Nivat 
studied infinitary rational relations in  \cite{Gire-Phd,Gire-Nivat}. 
These relations  are sets of pairs of infinite words  which are accepted 
by $2$-tape finite B\"uchi automata with asynchronous 
reading heads. 
The class of infinitary rational relations, which  extends both the 
class of finitary rational relations  and  the class of \orl s,  and 
the rational functions they may define, have  been much studied, 
see for example  \cite{ChoffrutGrigorieffG99,BCPS03,Simonnet92,Staiger97,PrieurPhd}. 
\nl Notice that a  rational relation $R\subseteq \Si_1^\om \times \Si_2^\om$ may be seen as 
an \ol~ over the alphabet $\Si_1 \times \Si_2$.  
\nl  A way to study the complexity of languages of infinite words  
accepted by finite machines is to study their topological complexity and firstly
to locate them with regard to 
the Borel and the projective hierarchies. 
This work is analysed 
for example in \cite{Staiger86a,Thomas90,eh,LescowThomas,Staiger97}.  
It is well known that every \ol~ accepted by a Turing machine with a 
B\"uchi or Muller acceptance condition is an analytic set and 
 that \orl s are boolean combinations of ${\bf \Pi}_2^0$-sets,  
hence ${\bf \Delta}_3^0$-sets \cite{Staiger97,PerrinPin}.  
\nl  The question of the topological complexity of  relations on infinite words also 
naturally arises and is asked by Simonnet in \cite{Simonnet92}. It is also posed in a more 
general form by Lescow and Thomas in \cite{LescowThomas} 
(for infinite labelled partial orders) and in \cite{Thomas88} 
where Thomas suggested to study reducibility notions and associated completeness results.  
\nl Every infinitary rational relation is an analytic set. 
We showed in \cite{Finkel03d} that there exist some infinitary rational relations 
which are analytic but non  Borel. Partial results about the Borel hierarchy of infinitary rational relations 
were first obtained in  \cite{Finkel03f,Fin07-Mold}, where  some 
${\bf \Si}^0_3$-complete and some ${\bf \Pi}^0_3$-complete infinitary rational relations were constructed.  In a recent conference paper 
we proved that for every non null recursive ordinal $\alpha$,  
there exist some  ${\bf \Si}^0_\alpha$-complete and some ${\bf \Pi}^0_\alpha$-complete infinitary rational relations,  \cite{Fin06b}.

\hs On the other hand,  William W. (Bill)  Wadge   studied    in \cite{Wadge83} a great refinement of the Borel hierarchy defined 
via  the notion of  reduction by continuous functions. 
He determined  the length of the new hierarchy obtained in that way, which is now called the Wadge hierarchy 
and has been much studied in descriptive set theory, 
see for example \cite{Moschovakis80,Kechris94,Duparc01}.
 It is then natural to ask for the Wadge hierarchy of classes of $\om$-languages accepted by 
finite machines, like B\"uchi automata or $2$-tape B\"uchi  automata. 
The Wagner hierarchy, 
effectively determined by Wagner, is actually, as noticed by Simonnet in \cite{Simonnet92},   
 the Wadge hierarchy of regular $\om$-languages;  
its length is the ordinal $\om^\om$ \cite{Wagner79,Selivanov95,Selivanov98}. 
The Wadge hierarchy of \ol s 
accepted by Muller {\it deterministic}  one blind (i. e. without zero-test)  counter 
 automata is an effective extension of the Wagner hierarchy studied in \cite{Fin01csl}. 
Wadge degrees of    {\it deterministic}             context free $\om$-languages have been determined by Duparc. The length of the Wadge hierarchy of 
{\it deterministic}             context free $\om$-languages 
 is the ordinal $\om^{(\om^2)}$ \cite{DFR,Duparc03}.
Selivanov has recently determined the Wadge hierarchy of $\om$-languages 
accepted by {\it deterministic} Turing machines; 
its length is $(\om_1^{\mathrm{CK}})^\om$    
\cite{Selivanov03a,Selivanov03b}.
\nl On the other hand, we recently proved in \cite{Fin-mscs06} that the Wadge  hierarchy of  $\om$-languages accepted by {\it non deterministic}
real time B\"uchi  $1$-counter automata is equal to the Wadge  hierarchy of  $\om$-languages accepted by {\it non deterministic} B\"uchi Turing machines. 
Using a simulation of real time $1$-counter automata, we show that, from a topological point of view,  
$2$-tape B\"uchi automata  have the same accepting power than 
Turing machines equipped with a  B\"uchi acceptance condition. 
The Borel and the Wadge hierarchies  of the class ${\bf RAT}_\om$ of infinitary rational relations accepted by $2$-tape B\"uchi automata are 
 equal to the Borel and the Wadge hierarchies  of 
$\om$-languages accepted  by real-time B\"uchi $1$-counter automata or by B\"uchi Turing machines. 
In particular,   for every non null recursive ordinal $\alpha$,  
there exist some  
${\bf \Si}^0_\alpha$-complete and some ${\bf \Pi}^0_\alpha$-complete infinitary rational relations.   And the supremum of the set of Borel ranks of 
infinitary rational relations is an ordinal $\gamma_2^1$ which is strictly greater than the first non recursive ordinal   
$\om_1^{\mathrm{CK}}$. 
This  very surprising result gives answers to 
questions of Simonnet \cite{Simonnet92} and of Lescow and Thomas \cite{Thomas90,LescowThomas}. 

 \hs This paper is an extended version of a conference paper  which appeared in the Proceedings 
of  the 23rd International Symposium on Theoretical Aspects of Computer Science, 
STACS 2006, \cite{Fin06b}. 

\hs  The paper is organized as follows. In section 2 we recall  
the notion of  2-tape  automata and of real time $1$-counter automata with B\"uchi acceptance condition. 
In section 3 we recall some definitions and properties of   Borel and Wadge hierarchies,  and we prove our main result  in section 4.

\section{$2$-tape automata and $1$-counter automata}

\noi We assume the reader to be familiar with the theory of formal ($\om$-)languages  
\cite{Thomas90,Staiger97}.
We shall use usual notations of formal language theory. 
\nl  When $\Si$ is a finite alphabet, a {\it non-empty finite word} over $\Si$ is any 
sequence $x=a_1\ldots a_k$, where $a_i\in\Sigma$ 
for $i=1,\ldots ,k$ , and  $k$ is an integer $\geq 1$. The {\it length}
 of $x$ is $k$, denoted by $|x|$.
 The {\it empty word} has no letter and is denoted by $\lambda$; its length is $0$. 
 For $x=a_1\ldots a_k$, we write $x(i)=a_i$  
and $x[i]=x(1)\ldots x(i)$ for $i\leq k$ and $x[0]=\lambda$.
 $\Sis$  is the {\it set of finite words} (including the empty word) over $\Sigma$.
 \nl  The {\it first infinite ordinal} is $\om$.
 An $\om$-{\it word} over $\Si$ is an $\om$ -sequence $a_1 \ldots a_n \ldots$, where for all 
integers $ i\geq 1$, ~
$a_i \in\Sigma$.  When $\sigma$ is an $\om$-word over $\Si$, we write
 $\sigma =\sigma(1)\sigma(2)\ldots \sigma(n) \ldots $,  where for all $i$,~ $\sigma(i)\in \Si$,
and $\sigma[n]=\sigma(1)\sigma(2)\ldots \sigma(n)$  for all $n\geq 1$ and $\sigma[0]=\lambda$.
\nl  
 The {\it set of } $\om$-{\it words} over  the alphabet $\Si$ is denoted by $\Si^\om$.
An  $\om$-{\it language} over an alphabet $\Sigma$ is a subset of  $\Si^\om$.  The complement (in $\Sio$) of an 
$\om$-language $V \subseteq \Sio$ is $\Sio - V$, denoted $V^-$.
 
\hs 
Infinitary rational relations 
are subsets of  $\Sio \times \Gao$,  where 
$\Si$ and  $\Ga$ are finite alphabets, which are accepted by 
$2$-tape B\"uchi automata  (2-BA). 

\begin{Deff}
A  2-tape B\"uchi automaton 
 is a sextuple $\mathcal{T}=(K, \Si, \Ga, \Delta, q_0, F)$, where 
$K$ is a finite set of states, $\Si$ and $\Ga$ are finite  alphabets, 
$\Delta$ is a finite subset of $K \times \Sis \times \Gas \times K$ called 
the set of transitions, $q_0$ is the initial state,  and $F \subseteq K$ is the set of 
accepting states. 
\nl A computation $\mathcal{C}$ of the  
2-tape B\"uchi automaton $\mathcal{T}$ is an infinite sequence of transitions 
$$(q_0, u_1, v_1, q_1), (q_1, u_2, v_2, q_2), \ldots ,(q_{i-1}, u_{i}, v_{i}, q_{i}), 
(q_i, u_{i+1}, v_{i+1}, q_{i+1}), \ldots $$
\noi The computation is said to be successful iff there exists a final state $q_f \in F$ 
and infinitely many integers $i\geq 0$ such that $q_i=q_f$. 
\nl The input word of the computation is $u=u_1.u_2.u_3 \ldots$
\nl The output word of the computation is $v=v_1.v_2.v_3 \ldots$
\nl Then the input and the output words may be finite or infinite. 
\nl The infinitary rational relation $R(\mathcal{T})\subseteq \Sio \times \Ga^\om$ 
accepted by the 2-tape B\"uchi automaton $\mathcal{T}$ 
is the set of pairs $(u, v) \in \Sio \times \Ga^\om$ such that $u$ and $v$ are the input 
and the output words of some successful computation $\mathcal{C}$ of $\mathcal{T}$. 
\nl The set of infinitary rational relations will be denoted by ${\bf RAT}_\om$. 
\end{Deff} 

\begin{Deff} 
A   ({\it real time})  $1$-counter machine  is a 4-tuple 
$\mathcal{M}$=$(K,\Si, \Delta, q_0)$,  where $K$ 
is a finite set of states, $\Sigma$ is a finite input alphabet, 
 $q_0\in K$ is the initial state, 
and the transition relation $\Delta$ is a subset of  
$K \times  \Si   \times \{0, 1\} \times K \times \{0, 1, -1\}$. 
\nl  
If  the machine $\mathcal{M}$ is in a state $q$ and 
$c \in \mathbb{N}$ is the content of the counter  then 
the  configuration (or global state)
 of $\mathcal{M}$ is  $(q, c)$.

\hs For $a\in \Si $, 
$q, q' \in K$ and $c \in \mathbb{N}$, if 
$(q, a, i, q', j) \in \Delta$,  where $i=0$ if $c=0$ 
and $i=1$ if $c\neq 0$,   then we write:
$$a: (q, c)\mapsto_{\mathcal{M}} (q', c+j)$$

\noi  Thus we see that the transition relation must satisfy:
 \nl if $(q, a,  i, q', j)   \in   \Delta$ and  $i=0$ then $j=0$ or $j=1$ (but $j$ may not be equal to $-1$).

\hs
Let $\sigma =a_1a_2 \ldots a_n $ be a finite word over $\Si$. 
A sequence of configurations $r=(q_i, c_{i})_{1\leq i \leq n+1}$  is called 
a  run of $\mathcal{M}$ on $\sigma$, starting in configuration 
$(p, c)$, iff:
\begin{enumerate}
\ite[(1)]  $(q_1, c_{1})=(p, c)$
\ite[(2)] 
 for each $i \in [1, n]$, 
 $a_i: (q_i, c_{i})\mapsto_{\mathcal{M}}
(q_{i+1},  c_{i+1})$ 
\end{enumerate}

\noi 
Let $\sigma =a_1a_2 \ldots a_n \ldots $ be an $\om$-word over $\Si$. 
An $\om$-sequence of configurations $r=(q_i, c_{i})_{i \geq 1}$ is called 
a run of $\mathcal{M}$ on $\sigma$, starting in configuration 
$(p, c)$, iff:
\begin{enumerate}
\ite[(1)]  $(q_1, c_{1})=(p, c)$

\ite[(2)]   for each $i\geq 1$, 
 $a_i: (q_i, c_{i})\mapsto_{\mathcal{M}}  
(q_{i+1},  c_{i+1})$  
\end{enumerate}
\noi
For every such run, $\mathrm{In}(r)$ is the set of all states entered infinitely
 often during the run $r$.
\nl
A  run $r$ of $M$ on $\sigma$, starting in configuration $(q_0, 0)$,
 will be simply called ``a run of $M$ on $\sigma$".
\end{Deff}

\begin{Deff} A ({\it real time}) B\"uchi $1$-counter automaton  is a 5-tuple 
\begin{center} 
$\mathcal{M}$=$(K,\Si, \Delta, q_0, F), $
\end{center} 
where $ \mathcal{M}'$=$(K,\Si, \Delta, q_0)$
is a ({\it real time}) $1$-counter machine and $F \subseteq K$ 
is the set of accepting  states.
The \ol~ accepted by $\mathcal{M}$ is 
\begin{center}
$L(\mathcal{M})$= $\{  \sigma\in\Si^\om \mid \mbox{  there exists a  run r
 of } \mathcal{M} \mbox{ on } \sigma \mbox{  such that } \mathrm{In}(r)
 \cap F \neq \emptyset \}$
\end{center}
\end{Deff}

\noi  The class of \ol s accepted by real time B\"uchi $1$-counter automata  will be 
denoted {\bf r}-${\bf BCL}(1)_\om$.

\section{Topology}

\subsection{Borel hierarchy and analytic sets}

\noi We assume the reader to be familiar with basic notions of topology which
may be found in \cite{Moschovakis80,LescowThomas,Kechris94,Staiger97,PerrinPin}.
There is a natural metric on the set $\Sio$ of  infinite words 
over a finite alphabet 
$\Si$ containing at least two letters which is called the {\it prefix metric} and defined as follows. For $u, v \in \Sio$ and 
$u\neq v$ let $\delta(u, v)=2^{-l_{\mathrm{pref}(u,v)}}$ where $l_{\mathrm{pref}(u,v)}$ 
 is the first integer $n$
such that the $(n+1)^{st}$ letter of $u$ is different from the $(n+1)^{st}$ letter of $v$. 
This metric induces on $\Sio$ the usual  Cantor topology for which {\it open subsets} of 
$\Sio$ are in the form $W.\Si^\om$, where $W\subseteq \Sis$.
A set $L\subseteq \Si^\om$ is a {\it closed set} iff its complement $\Si^\om - L$ 
is an open set.
Define now the {\it Borel Hierarchy} of subsets of $\Si^\om$:

\begin{Deff}
For a non-null countable ordinal $\alpha$, the classes ${\bf \Si}^0_\alpha$
 and ${\bf \Pi}^0_\alpha$ of the Borel Hierarchy on the topological space $\Si^\om$ 
are defined as follows:
\nl ${\bf \Si}^0_1$ is the class of open subsets of $\Si^\om$, 
 ${\bf \Pi}^0_1$ is the class of closed subsets of $\Si^\om$, 
\nl and for any countable ordinal $\alpha \geq 2$: 
\nl ${\bf \Si}^0_\alpha$ is the class of countable unions of subsets of $\Si^\om$ in 
$\bigcup_{\gamma <\alpha}{\bf \Pi}^0_\gamma$.
 \nl ${\bf \Pi}^0_\alpha$ is the class of countable intersections of subsets of $\Si^\om$ in 
$\bigcup_{\gamma <\alpha}{\bf \Si}^0_\gamma$.
\end{Deff}

\noi Recall some basic results about these classes:

\begin{Pro}
\noi  
\begin{enumerate}
\ite[(a)] ${\bf \Si}^0_\alpha \cup {\bf \Pi}^0_\alpha  \subsetneq  
{\bf \Si}^0_{\alpha +1}\cap {\bf \Pi}^0_{\alpha +1} $, for each countable 
ordinal  $\alpha \geq 1$. 
\ite[(b)] $\cup_{\gamma <\alpha}{\bf \Si}^0_\gamma = \cup_{\gamma <\alpha}{\bf \Pi}^0_\gamma 
\subsetneq {\bf \Si}^0_\alpha \cap {\bf \Pi}^0_\alpha $, for each countable limit ordinal 
$\alpha$. 
\ite[(c)] A set $W\subseteq X^\om$ is in the class ${\bf \Si}^0_\alpha $ iff its 
complement is in the class ${\bf \Pi}^0_\alpha $. 
\ite[(d)] ${\bf \Si}^0_\alpha  - {\bf \Pi}^0_\alpha  \neq \emptyset $ and 
${\bf \Pi}^0_\alpha  - {\bf \Si}^0_\alpha  \neq \emptyset $ hold 
 for every countable  ordinal $\alpha\geq 1$. 
\end{enumerate}
\end{Pro}

\noi For 
a countable ordinal $\alpha$,  a subset of $\Si^\om$ is a Borel set of {\it rank} $\alpha$ iff 
it is in ${\bf \Si}^0_{\alpha}\cup {\bf \Pi}^0_{\alpha}$ but not in 
$\bigcup_{\gamma <\alpha}({\bf \Si}^0_\gamma \cup {\bf \Pi}^0_\gamma)$.

\hs    
There are also some subsets of $\Si^\om$ which are not Borel. 
Indeed there exists another hierarchy beyond the Borel hierarchy, which is called the 
projective hierarchy and which is obtained from  the Borel hierarchy by 
successive applications of operations of projection and complementation.
The first level of the projective hierarchy is formed by the class of {\it analytic sets} and the class of {\it co-analytic sets} which are complements of 
analytic sets.  
In particular 
the class of Borel subsets of $\Si^\om$ is strictly included into 
the class  ${\bf \Si}^1_1$ of {\it analytic sets} which are 
obtained by projection of Borel sets. 

\begin{Deff} 
A subset $A$ of  $\Si^\om$ is in the class ${\bf \Si}^1_1$ of {\bf analytic} sets
iff there exists another finite set $Y$ and a Borel subset $B$  of  $(\Si \times Y)^\om$ 
such that $ x \in A \lra \exists y \in Y^\om $ such that $(x, y) \in B$, 
where $(x, y)$ is the infinite word over the alphabet $\Si \times Y$ such that
$(x, y)(i)=(x(i),y(i))$ for each  integer $i\geq 1$.
\end{Deff}

\noi  We now define completeness with regard to reduction by continuous functions. 
For a countable ordinal  $\alpha\geq 1$, a set $F\subseteq \Si^\om$ is said to be 
a ${\bf \Si}^0_\alpha$  
(respectively,  ${\bf \Pi}^0_\alpha$, ${\bf \Si}^1_1$)-{\it complete set} 
iff for any set $E\subseteq Y^\om$  (with $Y$ a finite alphabet): 
 $E\in {\bf \Si}^0_\alpha$ (respectively,  $E\in {\bf \Pi}^0_\alpha$,  $E\in {\bf \Si}^1_1$) 
iff there exists a continuous function $f: Y^\om \ra \Si^\om$ such that $E = f^{-1}(F)$. 
 ${\bf \Si}^0_n$
 (respectively ${\bf \Pi}^0_n$)-complete sets, with $n$ an integer $\geq 1$, 
 are thoroughly characterized in \cite{Staiger86a}.  

\hs In particular  $\mathcal{R}=(0^\star.1)^\om$  
is a well known example of 
${\bf \Pi}^0_2 $-complete subset of $\{0, 1\}^\om$. It is the set of 
$\om$-words over $\{0, 1\}$ having infinitely many occurrences of the letter $1$. 
Its  complement 
$\{0, 1\}^\om - (0^\star.1)^\om$ is a 
${\bf \Si}^0_2 $-complete subset of $\{0, 1\}^\om$.

\hs We recall now the definition of the  arithmetical hierarchy of  \ol s which form the effective analogue to the 
hierarchy of Borel sets of finite ranks. 
\nl Let $X$ be a finite alphabet. An \ol~ $L\subseteq X^\om$  belongs to the class 
$\Si_n$ if and only if there exists a recursive relation 
$R_L\subseteq (\mathbb{N})^{n-1}\times X^\star$  such that
$$L = \{\sigma \in X^\om \mid \exists a_1\ldots Q_na_n  \quad (a_1,\ldots , a_{n-1}, 
\sigma[a_n+1])\in R_L \}$$

\noi where $Q_i$ is one of the quantifiers $\fa$ or $\exists$ 
(not necessarily in an alternating order). An \ol~ $L\subseteq X^\om$  belongs to the class 
$\Pi_n$ if and only if its complement $X^\om - L$  belongs to the class 
$\Si_n$.  The inclusion relations that hold  between the classes $\Si_n$ and $\Pi_n$ are 
the same as for the corresponding classes of the Borel hierarchy. 
 The classes $\Si_n$ and $\Pi_n$ are  included in the respective classes 
${\bf \Si_n^0}$ and ${\bf \Si_n^0}$ of the Borel hierarchy, and cardinality arguments suffice to show that these inclusions are strict. 

\hs  As in the case of the Borel hierarchy, projections of arithmetical sets 
(of the second $\Pi$-class) lead 
beyond the arithmetical hierarchy, to the analytical hierarchy of \ol s. The first class 
of this hierarchy is the (lightface) class $\Si^1_1$ of {\it effective analytic sets} 
 which are obtained by projection of arithmetical sets.
 The (lightface)  class $\Pi_1^1$ of  {\it effective co-analytic sets} 
 is simply the class of complements of effective analytic sets. We denote as usual $\Delta_1^1 = \Si^1_1 \cap \Pi_1^1$. 

\hs  It is well known  that an \ol~ $L\subseteq X^\om$ is in the class $\Si_1^1$
iff it is accepted by a non deterministic Turing machine 
with a   B\"uchi or Muller acceptance condition  \cite{Staiger97}. 

\hs   Borel ranks of  (lightface) $\Delta_1^1$ sets   are the (recursive) 
ordinals  $\gamma < \om_1^{\mathrm{CK}}$, where $ \om_1^{\mathrm{CK}}$
 is the first non-recursive ordinal, usually called the Church-Kleene ordinal.  
Moreover, for every non null  ordinal $\alpha < \om_1^{\mathrm{CK}}$, there exist some  
${\bf \Si}^0_\alpha$-complete and some  ${\bf \Pi}^0_\alpha$-complete sets in the class $\Delta_1^1$. 
  On the other hand,  Kechris, Marker and Sami proved in \cite{KMS89} that the supremum 
of the set of Borel ranks of  (lightface) $\Si_1^1$-sets is an ordinal $\gamma_2^1$ which 
 is strictly greater than the ordinal $\delta_2^1$ which is the first non $\Delta_2^1$ ordinal. 
Thus  the ordinal  $\gamma_2^1$ is also strictly greater than the Church-Kleene ordinal
$ \om_1^{\mathrm{CK}}$.  
The exact value of the ordinal $\gamma_2^1$ may depend on axioms of  set theory  \cite{KMS89}. It is consistent with the axiomatic system 
{\bf ZFC } that $\gamma_2^1$ is equal to the ordinal $\delta_3^1$ which is the first non $\Delta_3^1$ ordinal  
(because $\gamma_2^1= \delta_3^1$  in {\bf ZFC + (V=L)}). On the other hand  the axiom of 
$\Pi_1^1$-determinacy implies that $\gamma_2^1 < \delta_3^1$.  For more details,   we refer the reader to  \cite{KMS89} and to 
a textbook of set theory like \cite{Jech}. 
\nl Notice however that it seems still unknown  whether {\it every } non null ordinal $\gamma < \gamma_2^1$ is the Borel rank 
of a (lightface)  $\Si_1^1$-set.

\subsection{Wadge hierarchy}

\noi We now introduce the Wadge hierarchy, which is a great refinement of the Borel hierarchy defined 
via reductions by continuous functions, \cite{Wadge83,Duparc01}. 

\begin{Deff}[Wadge \cite{Wadge83}] Let $X$, $Y$ be two finite alphabets. 
For $L\subseteq X^\om$ and $L'\subseteq Y^\om$, $L$ is said to be Wadge reducible to $L'$
($L\leq _W L')$ iff there exists a continuous function $f: X^\om \ra Y^\om$, such that
$L=f^{-1}(L')$.
\nl $L$ and $L'$ are Wadge equivalent iff $L\leq _W L'$ and $L'\leq _W L$. 
This will be denoted by $L\equiv_W L'$. And we shall say that 
$L<_W L'$ iff $L\leq _W L'$ but not $L'\leq _W L$.
\nl  A set $L\subseteq X^\om$ is said to be self dual iff  $L\equiv_W L^-$, and otherwise 
it is said to be non self dual.
\end{Deff}

\noi
 The relation $\leq _W $  is reflexive and transitive,
 and $\equiv_W $ is an equivalence relation.
\nl The {\it equivalence classes} of $\equiv_W $ are called {\it Wadge degrees}. 
\nl The Wadge hierarchy $WH$ is the class of Borel subsets of a set  $X^\om$, where  $X$ is a finite set,
 equipped with $\leq _W $ and with $\equiv_W $.
\nl  For $L\subseteq X^\om$ and $L'\subseteq Y^\om$, if   
$L\leq _W L'$ and $L=f^{-1}(L')$  where $f$ is a continuous 
function from $ X^\om$  into $Y^\om$, then $f$ is called a continuous reduction of $L$ to 
$L'$. Intuitively it means that $L$ is less complicated than $L'$ because 
to check whether $x\in L$ it suffices to check whether $f(x)\in L'$ where $f$ 
is a continuous function. Hence the Wadge degree of an \ol~
is a measure 
of its topological complexity. 

\hs 
Notice  that in the above definition, we consider that a subset $L\subseteq  X^\om$ is given
together with the alphabet $X$. This is important as it is shown by the following simple example. 
Let $L_1=\{0, 1\}^\om \subseteq \{0, 1\}^\om$ and $L_2=\{0, 1\}^\om \subseteq \{0, 1, 2\}^\om$. So the languages $L_1$ and $L_2$ are equal 
but considered over the different alphabets  $X_1=\{0, 1\}$ and $X_2=\{0, 1, 2\}$. It turns out that $L_1 <_W L_2$. In fact $L_1$ is open {\it and } 
closed in $X_1^\om$ while $L_2$ is closed but non open in $X_2^\om$. 

\hs  We can now define the {\it Wadge class} of a set $L$:

\begin{Deff}
Let $L$ be a subset of $X^\om$. The Wadge class of $L$ is :
$$[L]= \{ L' \mid  L'\subseteq Y^\om \mbox{ for a finite alphabet }Y   \mbox{  and  } L'\leq _W L \}.$$ 
\end{Deff}

\noi Recall that each {\it Borel class} ${\bf \Si}^0_\alpha$ and ${\bf \Pi}^0_\alpha$ 
is a {\it Wadge class}. 
\nl A set $L\subseteq X^\om$ is a ${\bf \Si}^0_\alpha$
 (respectively ${\bf \Pi}^0_\alpha$)-{\it complete set} iff for any set 
$L'\subseteq Y^\om$, $L'$ is in 
${\bf \Si}^0_\alpha$ (respectively ${\bf \Pi}^0_\alpha$) iff $L'\leq _W L $ .
  It follows from the study of the Wadge hierarchy that a set $L\subseteq X^\om$ is a ${\bf \Si}^0_\alpha$
 (respectively, ${\bf \Pi}^0_\alpha$)-{\it complete set} iff it is in ${\bf \Si}^0_\alpha$ but not in ${\bf \Pi}^0_\alpha$
 (respectively, in ${\bf \Pi}^0_\alpha$ but not in ${\bf \Si}^0_\alpha$).

\hs  There is a close relationship between Wadge reducibility
 and games which we now introduce.  

\begin{Deff}[Wadge \cite{Wadge83}]  Let 
$L\subseteq X^\om$ and $L'\subseteq Y^\om$. 
The Wadge game  $W(L, L')$ is a game with perfect information between two players,
player 1 who is in charge of $L$ and player 2 who is in charge of $L'$.
\nl Player 1 first writes a letter $a_1\in X$, then player 2 writes a letter
$b_1\in Y$, then player 1 writes a letter $a_2\in  X$, and so on. 
\nl The two players alternatively write letters $a_n$ of $X$ for player 1 and $b_n$ of $Y$
for player 2.
\nl After $\om$ steps, the player 1 has written an $\om$-word $a\in X^\om$ and the player 2
has written an $\om$-word $b\in Y^\om$.
 The player 2 is allowed to skip, even infinitely often, provided he really writes an
$\om$-word in  $\om$ steps.
\nl The player 2 wins the play iff [$a\in L \lra b\in L'$], i.e. iff : 
\begin{center}
  [($a\in L ~{\rm and} ~ b\in L'$)~ {\rm or} ~ 
($a\notin L ~{\rm and}~ b\notin L'~{\rm and} ~ b~{\rm is~infinite}  $)].
\end{center}
\end{Deff}

\noi
Recall that a strategy for player 1 is a function 
$\sigma :(Y\cup \{s\})^\star\ra X$.
And a strategy for player 2 is a function $f:X^+\ra Y\cup\{ s\}$.
\nl $\sigma$ is a winning stategy  for player 1 iff he always wins a play when
 he uses the strategy $\sigma$, i.e. when the  $n^{th}$  letter he writes is given
by $a_n=\sigma (b_1\ldots b_{n-1})$, where $b_i$ is the letter written by player 2 
at step $i$ and $b_i=s$ if player 2 skips at step $i$.
\nl A winning strategy for player 2 is defined in a similar manner.

\hs   Martin's Theorem states that every Gale-Stewart Game $G(X)$,  with $X$ a Borel set, 
is determined,  see \cite{Kechris94}. This implies the following determinacy result :

\begin{theorem} [Wadge] Let $L\subseteq X^\om$ and $L'\subseteq Y^\om$ be two Borel sets, where
$X$ and $Y$ are finite  alphabets. Then the Wadge game $W(L, L')$ is determined :
one of the two players has a winning strategy. And $L\leq_W L'$ iff the player 2 has a 
winning strategy  in the game $W(L, L')$.
\end{theorem}

\begin{theorem} [Wadge]\label{wh}
Up to the complement and $\equiv _W$, the class of Borel subsets of $X^\om$,
 for  a finite alphabet $X$, is a well ordered hierarchy.
 There is an ordinal $|WH|$, called the length of the hierarchy, and a map
$d_W^0$ from $WH$ onto $|WH|-\{0\}$, such that for all $L, L' \subseteq X^\om$:
\nl $d_W^0 L < d_W^0 L' \lra L<_W L' $  and 
\nl $d_W^0 L = d_W^0 L' \lra [ L\equiv_W L' $ or $L\equiv_W L'^-]$.
\end{theorem}

\noi 
 The Wadge hierarchy of Borel sets of {\it finite rank }
has  length $^1\varepsilon_0$ where $^1\varepsilon_0$
 is the limit of the ordinals $\alpha_n$ defined by $\alpha_1=\om_1$ and 
$\alpha_{n+1}=\om_1^{\alpha_n}$ for $n$ a non negative integer, $\om_1$
 being the first non countable ordinal. Then $^1\varepsilon_0$ is the first fixed 
point of the ordinal exponentiation of base $\om_1$. The length of the Wadge hierarchy 
of Borel sets in ${\bf \Delta^0_\om}= {\bf \Si^0_\om}\cap {\bf \Pi^0_\om}$ 
  is the $\om_1^{th}$ fixed point 
of the ordinal exponentiation of base $\om_1$, which is a much larger ordinal. The length 
of the whole Wadge hierarchy of Borel sets is a huge ordinal, with regard 
to the $\om_1^{th}$ fixed point 
of the ordinal exponentiation of base $\om_1$. It has been determined by Wadge and is described in \cite{Wadge83,Duparc01} 
by the use of the Veblen functions.

\section{Wadge hierarchy of  infinitary rational relations}

\hs We have proved in \cite{Fin-mscs06} the following result.

\begin{theorem}[\cite{Fin-mscs06}]\label{thewad}    
The Wadge hierarchy of the class 
{\bf r}-${\bf BCL}(1)_\om$ 
is equal to the Wadge hierarchy of the class 
 $\Sigma^1_1$ of $\om$-languages accepted by Turing machines with a B\"uchi acceptance 
condition. 
\end{theorem}

\noi We are going to prove a similar  result for the class ${\bf RAT}_\om$, 
 using a simulation of  $1$-counter automata. 

\begin{theorem}\label{thewad2}    
The Wadge hierarchy of the class 
${\bf RAT}_\om$ 
is equal to the Wadge hierarchy of the class {\bf r}-${\bf BCL}(1)_\om$ hence also  of the class 
 $\Sigma^1_1$ of $\om$-languages accepted by Turing machines with a B\"uchi acceptance 
condition. 
\end{theorem}

\noi We now first  define a coding of  an $\om$-word over a finite alphabet $\Si$, such that $0\in \Si$, 
by an  $\om$-word over the  alphabet $\Ga = \Si \cup \{A\}$, where  $A$ is an additional letter 
not in $\Si$. 

\hs For $x\in \Sio$  the $\om$-word $h(x)$ is defined by: 
$$h(x) = A.0.x(1).A.0^2.x(2).A.0^3.x(3).A.0^4.x(4).A. \ldots .A.0^n.x(n).A.0^{n+1}.x(n+1).A \ldots$$
\noi Then it is easy to see that the mapping $h$ from $\Sio$ into $(\Si \cup \{A\})^\om$ is continuous and injective.

\hs Let now  $\alpha$ be the $\om$-word over the alphabet 
$\Si \cup\{A\}$ which is defined by:

$$\alpha = A.0.A.0^2.A.0^3.A.0^4.A.0^5.A \ldots A.0^n.A.0^{n+1}.A \ldots$$

\noi We can now state the following Lemma.

\begin{Lem}\label{R1}
Let $\Si$  be a finite alphabet such that $0\in \Si$, 
$\alpha$ be the $\om$-word over  $\Si \cup\{A\}$ defined as above, and 
 $L \subseteq \Sio$ be in  {\bf r}-${\bf BCL}(1)_\om$.
Then there exists  an infinitary rational relation 
$R_1 \subseteq (\Si\cup\{A\})^\om \times (\Si \cup\{A\})^\om$ such that:
$$\fa x\in \Si^{\om}~~~ (x\in L) \mbox{  iff } ( (h(x), \alpha) \in R_1 )$$ 
\end{Lem}

\proo Let $\Si$  be a finite alphabet such that $0\in \Si$, $\alpha$ be the $\om$-word 
over  $\Si \cup\{A\}$ defined as above, and  $L = L(\mathcal{A}) \subseteq \Sio$, where 
$\mathcal{A}$=$(K,\Si, \Delta, q_0, F)$ is a real time $1$-counter B\"uchi automaton. 

\hs We define now the relation $R_1$.
A pair 
$y=(y_1, y_2)$ of $\om$-words over the alphabet $\Si\cup\{A\}$ is in $R_1$ 
if and only if it is in the form

\hs $y_1 = A.u_1.v_1.x(1).A.u_2.v_2.x(2).A.u_3.v_3.x(3).A  
\ldots A.u_{n}.v_{n}.x(n).A. \ldots$
\nl $y_2 = A.w_1.z_1.A.w_2.z_2.A.w_3.z_3.A \ldots 
 A.w_{n}.z_{n}.A \ldots$

\hs where $|v_1|=0$ and 
for all integers $i\geq 1$, 
$$ u_i ,v_i, w_i, z_i \in 0^\star \mbox{ and } x(i) \in \Si  \mbox{  and   }  $$ 
$$~~~~~ |u_{i+1}|=|z_i|+1 $$ 

\noi and there is a sequence $(q_i)_{i\geq 0}$  of states of $K$ 
 such that  for all integers 
$i\geq 1$:  

$$  x(i) : ( q_{i-1}, |v_i| ) \mapsto_{\mathcal{A}} 
(q_i, |w_i| )$$

\noi Moreover some state $q_f \in F$ occurs infinitely often in the sequence $(q_i)_{i\geq 0}$. 
\nl  Notice that the state $q_0$ of the sequence $(q_i)_{i\geq 0}$  is also the initial state 
of $\mathcal{A}$. 

\hs Notice that the main idea is that we try to simulate, using a $2$-tape automaton,  the reading of the infinite word $x(1).x(2).x(3) \ldots$ by  the 
real time $1$-counter B\"uchi automaton $\mathcal{A}$. The initial value of the counter is $|v_1|$ and the value  of the counter 
after the reading of the letter $x(1)$ by  $\mathcal{A}$  is $|w_1|$ which is  on the second tape. Now the $2$-tape automaton accepting 
$R_1$ would need to read again the value $|w_1|$ 
in order to compare it to the value of the counter after the reading of $x(2)$ by the $1$-counter  automaton $\mathcal{A}$. 
This is not directly possible so the simulation does not work on every pair of $R_1$. However,  using the very special shape of  pairs in 
$h(\Sio) \times \{\alpha\}$, the simulation will be possible on a pair $(h(x), \alpha)$. Then for such a pair  $ (h(x), \alpha) \in R_1$ written 
in the above form $(y_1, y_2)$,  we have $|v_2|=|w_1|$ and then  the simulation can continue from the value $|v_2|$ of the counter, and so on.

\hs We now give the details of the proof. 
\nl Let  $x\in \Si^{\om}$ be  such that  $(h(x), \alpha) \in R_1$. We are going to prove that $x\in L$. 

\hs  By hypothesis  $ (h(x), \alpha) \in R_1$ thus there are finite words  $u_i ,v_i, w_i, z_i \in 0^\star$ such that 
 $|v_1|=0$ and for all integers $i\geq 1$, $|u_{i+1}|=|z_i|+1 $, and  

\hs $h(x) = A.u_1.v_1.x(1).A.u_2.v_2.x(2).A.u_3.v_3.x(3).A  
\ldots A.u_{n}.v_{n}.x(n).A. \ldots$

\hs $\alpha  = A.w_1.z_1.A.w_2.z_2.A.w_3.z_3.A \ldots 
 A.w_{n}.z_{n}.A \ldots$

\hs Moreover  there is a sequence $(q_i)_{i\geq 0}$  of states of $K$ 
 such that  for all integers 
$i\geq 1$:  

$$  x(i) : ( q_{i-1}, |v_i| ) \mapsto_{\mathcal{A}} 
(q_i, |w_i| )$$

\noi and  some state $q_f \in F$ occurs infinitely often in the sequence $(q_i)_{i\geq 0}$. 

\hs 
On the other side we have: 
\nl  $h(x) = A.0.x(1).A.0^2.x(2).A.0^3.x(3).A \ldots A.0^n.x(n).A.0^{n+1}.x(n+1).A \ldots$
\nl $\alpha =A.0.A.0^2.A.0^3.A.0^4.A  \ldots A.0^n.A \ldots$

\hs So we have $|u_1.v_1|=1$ and $|v_1|=0$ and $x(1): ( q_{0}, |v_1| ) \mapsto_{\mathcal{A}} 
(q_1, |w_1| )$. But $|w_1.z_1|=1$,  $|u_2.v_2|=2$, and $|u_2|=|z_1|+1$ thus $|v_2|=|w_1|$. 

\hs We are going to prove  in a similar way  that for all integers $i\geq 1$ it holds that $|v_{i+1}|=|w_i|$. 
\nl We know that $|w_i.z_i|=i$, $|u_{i+1}.v_{i+1}|=i+1$, and $|u_{i+1}|=|z_i|+1$ thus $|w_i|=|v_{i+1}|$. 

\hs Then for all $i\geq 1$,  $x(i) : ( q_{i-1}, |v_i| ) \mapsto_{\mathcal{A}} (q_i, |v_{i+1}| )$. 
\nl So if we set $c_i=|v_i|$, $(q_{i-1}, c_{i})_{i\geq 1}$ is an accepting run of $\mathcal{A}$ on $x$  and this implies that  
$x\in L$. 
\nl Conversely it is easy to prove that if $x\in L$ then $(h(x), \alpha)$ may be written in the form of $(y_1, y_2)\in R_1$. 

\hs It remains  to prove that the above defined relation $R_1$ is an infinitary rational 
relation.  It is easy to find a $2$-tape  B\"uchi automaton $\mathcal{T}$ accepting  
the  relation $R_1$.            

    \ep

\begin{Lem}\label{complement} The set 
$$R_2 = (\Si\cup\{A\})^\om\times (\Si \cup\{A\})^\om - ( h(\Si^{\om}) \times \{\alpha\} )$$
\noi is an infinitary rational relation.  
\end{Lem}

\proo By definition of the mapping $h$, we know that a pair of $\om$-words 
over   the alphabet  $(\Si\cup\{A\})$ is in $h(\Si^{\om}) \times \{\alpha\}$ iff 
 it is  in the form $(\sigma_1, \sigma_2)$, 
where
\hs  $\sigma_1 = A.0.x(1).A.0^2.x(2).A.0^3.x(3).A \ldots 
.A.0^n.x(n).A.0^{n+1}.x(n+1).A \ldots$
\nl  $\sigma_2 = \alpha = A.0.A.0^2.A.0^3.A \ldots A.0^n.A.0^{n+1}.A \ldots$

\hs where for all integers $i\geq 1$,  $x(i)\in \Si$. 

\hs So it is easy to see that 
$(\Si\cup\{A\})^\om\times (\Si \cup\{A\})^\om - ( h(\Si^{\om}) \times \{\alpha\} )$
is the union of the sets $\mathcal{C}_j$ where:

\begin{itemize} 

\ite $\mathcal{C}_1$ is formed by pairs  $(\sigma_1, \sigma_2)$ where 
\nl $\sigma_1$ has not any initial segment in $A.\Si^2.A.\Si^3.A$, 
or $\sigma_2$ has not any initial segment in $A.\Si.A.\Si^2.A$.

\ite $\mathcal{C}_2$ is formed by pairs  $(\sigma_1, \sigma_2)$ where 
\nl $\sigma_2 \notin  (A.0^+)^\om$, or $\sigma_1 \notin (A.0^+.\Si)^\om$.  

\ite $\mathcal{C}_3$ is formed by pairs  $(\sigma_1, \sigma_2)$ where 
\nl $\sigma_1 = A.w_1.A.w_2.A.w_3.A \ldots A.w_n.A.u.A.z_1 $
\nl $\sigma_2 = A.w'_1.A.w'_2.A.w'_3.A \ldots A.w'_n.A.v.A.z_2 $

\hs where $n$ is an integer $\geq 1$,  for all $i \leq n$~  $w_i, w'_i \in \Sis$,  
$z_1, z_2 \in (\Si\cup\{A\})^\om$ and 
$$u, v \in \Sis \mbox{  and }  |u| \neq |v| + 1$$

\ite $\mathcal{C}_4$ is formed by pairs  $(\sigma_1, \sigma_2)$ where 
\nl $\sigma_1 = A.w_1.A.w_2.A.w_3.A.w_4 \ldots A.w_n.A.w_{n+1}.A.v.A.z_1 $
\nl $\sigma_2 = A.w'_1.A.w'_2.A.w'_3.A.w'_4 \ldots A.w'_n.A.u.A.z_2 $

\hs where $n$ is an integer $\geq 1$,  for all $i \leq n$~  $w_i, w'_i \in \Sis$, 
$w_{n+1} \in \Sis$, 
$z_1, z_2 \in (\Si\cup\{A\})^\om$ and 
$$u, v \in \Sis \mbox{  and }  |v|\neq |u|+2$$

\end{itemize}

\noi Each set $\mathcal{C}_j$, $1\leq j\leq 4$, is easily seen to be an infinitary 
rational relation $\subseteq (\Si\cup\{A\})^\om \times (\Si\cup\{A\})^\om$ (the detailed 
proof is left to the reader).  The class ${\bf RAT}_\om$ is closed under finite union thus

$$R_2 = (\Si\cup\{A\})^\om\times (\Si \cup\{A\})^\om - ( h(\Si^{\om}) \times \{\alpha\} )
 = \bigcup_{1\leq j\leq 4} \mathcal{C}_j$$

\noi is an infinitary rational relation. \ep  

\hs  As in  \cite{Fin-mscs06},  we  are going to consider  first non self dual sets to prove Theorem \ref{thewad2}. 
We recall the definition of Wadge degrees 
introduced by Duparc in \cite{Duparc01} and which is a slight modification of the previous one. 

\begin{Deff}
\noi
\begin{enumerate}
\ite[(a)]  $d_w(\emptyset)=d_w(\emptyset^-)=1$
\ite[(b)]  $d_w(L)=sup \{d_w(L')+1 ~\mid ~ L' {\rm ~non~ self ~dual~ and~}
L'<_W L \} $
\nl (for either $L$ self dual or not, $L>_W \emptyset).$
\end{enumerate}
\end{Deff}

\noi  We are going now to
introduce the operation of sum of 
sets of infinite words which has as  
counterpart the ordinal
addition  over Wadge degrees.

\begin{Deff}[Wadge, see \cite{Wadge83,Duparc01}]
Assume that $X\subseteq Y$ are two finite alphabets,
  $Y-X$ containing at least two elements, and that
$\{X_+, X_-\}$ is a partition of $Y-X$ in two non empty sets.
 Let $L \subseteq X^{\om}$ and $L' \subseteq Y^{\om}$, then
 $$L' + L =_{df} L\cup \{ u.a.\beta  ~\mid  ~ u\in X^\star , ~(a\in X_+
~and ~\beta \in L' )~
or ~(a\in X_- ~and ~\beta \in L'^- )\}$$
\end{Deff}

\noi This operation is closely related to the {\it ordinal sum}
 as it is stated in the following:

\begin{theorem}[Wadge, see \cite{Wadge83,Duparc01}]\label{thesum}
Let $X\subseteq Y$, $Y-X$ containing at least two elements,
   $L \subseteq X^{\om}$ and $L' \subseteq Y^{\om}$ be 
non self dual  Borel sets.
Then $(L+L')$ is a non self dual Borel set and
$d_w( L'+L )= d_w( L' ) + d_w( L )$.
\end{theorem}

\noi A player in charge of a set $L'+L$ in a Wadge game is like a player in charge of the set $L$ but who 
can, at any step of the play,    erase  his previous play and choose to be this time in charge of  $L'$ or of $L'^-$. 
Notice that he can do this only one time during a play. 
 The following lemma was proved in \cite{Fin-mscs06}. 

\begin{Lem}\label{6-10}
Let $L \subseteq \Sio$ be a non self dual  Borel set such that $d_w( L )\geq \om$. Then it holds that $L \equiv_W \emptyset + L$. 
\end{Lem}

\noi Notice that in the above lemma, $\emptyset$ is viewed as the empty set over an alphabet $\Si'$ such that 
$\Si \subseteq \Si'$ and cardinal ($\Si' - \Si$) $\geq 2$. 

\hs 
\proo 
Assume that  $L \subseteq \Sio$ is a non self dual  Borel set and  that $d_w( L )\geq \om$.
We know that $\emptyset$ is a non self dual Borel set and that $d_w( \emptyset ) = 1$. Thus, by Theorem \ref{thesum}, 
it holds that $d_w(  \emptyset + L ) = d_w(  \emptyset ) + d_w( L ) = 1 + d_w( L ) $. But by hypothesis $d_w( L )\geq \om$ and this implies 
that  $1 + d_w( L ) = d_w( L )$. So we have proved that $d_w(  \emptyset + L ) =  d_w( L )$. 
 \nl On the other hand $L$ is non self dual and $d_w(  \emptyset + L ) =  d_w( L )$ imply that only two cases may happen : 
either $\emptyset + L  \equiv_W  L$  or  $\emptyset + L \equiv_W L^-$. 
\nl But it is easy to see that $L \leq_W \emptyset + L$. For that purpose consider the Wadge game $W( L, \emptyset + L)$. 
Player 2 has clearly a winning strategy which consists in copying the play of Player 1 thus $L \leq_W \emptyset + L$. 
This implies that $\emptyset + L \equiv_W L^-$ cannot hold so $\emptyset + L  \equiv_W  L$. 
\ep

\hs We can now state the following lemma.

\begin{Lem}\label{6-11}
Let  $L \subseteq \Sio$ be a non self dual  Borel set accepted by a real time  B\"uchi  $1$-counter automaton. 
Then there is an infinitary rational relation $R \subseteq   (\Si\cup\{A\})^\om\times (\Si \cup\{A\})^\om$ such that $L \equiv_W R$.  
\end{Lem}

\proo 
It is well known that there are regular $\om$-languages of every finite Wadge degree, \cite{Staiger97,Selivanov98}. These $\om$-languages 
are Boolean combinations of open sets.  So we have only to consider the case of non self dual Borel sets of Wadge 
degrees greater than or equal to $\om$. 

\hs   Let  then $L = L(\mathcal{A})\subseteq \Sio$ be a non self dual Borel set
accepted by a (real time)  B\"uchi $1$-counter automaton $\mathcal{A}$ such that $d_w( L )\geq \om$. 

\hs Let $\Ga= \Si \cup \{A\}$ and $R_1 \subseteq \Gao \times \Gao$ be the infinitary rational relation constructed from $L(\mathcal{A})$ 
as in the proof of Lemma \ref{R1} and let 

$$R = R_1 \cup R_2 \subseteq \Gao \times \Gao$$
\noi The class ${\bf RAT}_\om$ is closed under finite union therefore $R$ is an 
infinitary rational relation. 

\hs Lemma \ref{R1} and the definition of $R_2$ imply  that 
$R_\alpha = \{\sigma \in \Ga^\om \mid (\sigma, \alpha) \in R \}$ is equal to the set 
$\mathcal{L}= h(L)  \cup (h(\Si^{\om}))^- $.

\hs Moreover, for all $u\in \Ga^\om-\{\alpha\}$, 
$R_u = \{\sigma \in \Ga^\om \mid (\sigma, u) \in R \} = \Ga^\om$ holds by 
definition of $R_2$.       

\hs  Then  
 $R$ may be written as the  union: 
 $$R = \mathcal{L} \times \{\alpha\} ~~ \bigcup  ~~\Gao \times (\Gao - \{\alpha\})$$
\noi or 
$$R =  h(L)  \times \{\alpha\} ~~ \bigcup  ~~(h(\Si^{\om}))^- \times \Gao  ~~ \bigcup  ~~\Gao \times (\Gao - \{\alpha\})$$

\noi 
 It is easy to see that  $(h(\Si^{\om}))^- \times \Gao$ and $\Gao \times (\Gao - \{\alpha\})$  are  open  
subsets of $\Gao \times \Gao$. Thus $(h(\Si^{\om}))^- \times \Gao  ~~ \bigcup  ~~\Gao \times (\Gao - \{\alpha\})$ is an open subset 
of $\Gao \times \Gao$. We denote $R'$ this open set so we have $R =  h(L)  \times \{\alpha\} ~~ \bigcup  ~~R'$. 

\hs In order to prove that $R \leq_W L$ it suffices to prove that $ R \leq_W \emptyset + L $ because 
Lemma \ref{6-10} states that $ \emptyset + L \equiv_W  L$.  We consider the Wadge game $W(R, \emptyset + L)$. Player 1 is in charge 
of the set $R \subseteq \Gao \times \Gao$ and 
Player 2 is in charge of the set $\emptyset + L $. 

\hs Player 2 has a winning strategy in this game which we now describe. Whenever Player 1  ``remains" in the closed set $h(\Sio)  \times \{\alpha\}$, i.e. 
whenever the word written by Player 1 is a prefix of some $\om$-word in $h(\Sio)  \times \{\alpha\}$, then Player 2 follows the play of Player 1 but skipping often
in such a way that he has written the word $x(1).x(2).x(3) \ldots x(n)$ when player 1 has written the word 
$$(A.0.x(1).A.0^2.x(2).A.0^3.x(3).A \ldots A.0^n.x(n), ~\alpha[3+4+5+ \ldots + n+2]).$$
\noi If  Player 1  ``remains" forever 
in the closed set $h(\Sio)  \times \{\alpha\}$  then after $\om$ steps 
Player 1 has written an $\om$-word $(h(x), \alpha)$ for some $x \in \Sio$, and Player 2 has written $x$. So in that case 
$(h(x), \alpha)$ is in $R$ iff   $x$ is in $L$ iff  $x$ is in $\emptyset + L $.  

\hs But if at some step of the play, Player 1 ``goes out of" the closed set  $h(\Sio)  \times \{\alpha\}$, 
 because the word he has now 
written is not a prefix of any $\om$-word of   $h(\Sio)  \times \{\alpha\}$ then its final word will be surely outside     $h(\Sio)  \times \{\alpha\}$ 
hence it will be surely  in $R$. 
Player 2 can now write a letter of $\Si' -\Si$ in such a way that he is now like a player in charge of 
 $(\Si')^\om$   and he can 
now write an $\om$-word $u \in (\Si')^\om$ so that his final $\om$-word will be in $\emptyset + L $. Thus Player 2 wins this play too. 

\hs We have then proved that $R \leq_W L$. 
 
\hs In order to prove that $L \leq_W R$ we consider the function 
  $g: \Si^{\om} \ra 
 (\Si\cup\{A\})^\om \times (\Si\cup\{A\})^\om $   defined  by: ~~~~~~~
$\fa x \in \Si^{\om}~~~~~~~ g(x) = (h(x) , \alpha)$.

\noi  It is easy to see that $g$ is continuous because $h$ is continuous. By construction 
it turns out that 
for all $\om$-words $x\in \Si^{\om}$, ~~~$(x\in L)$ iff $( (h(x) , \alpha)\in R )$ iff  $(g(x)\in R)$. 
This means that $g^{-1}(R)= L$. This implies that $L \leq_W R$.   

\hs Finally we have proved that  $R \leq_W L \leq_W R$, so  the infinitary rational relation $R$ is Wadge equivalent to the $\om$-language $L$ 
and this ends the proof.            \ep

\hs {\bf End of Proof of Theorem \ref{thewad2}. }   
Let  $L \subseteq \Sio$ be a   Borel set acccepted by a real time B\"uchi $1$-counter automaton $\mathcal{A}$. 
If  $L$ is non self dual,  then by Lemma   \ref{6-11} 
there is an infinitary rational relation $R \subseteq   (\Si\cup\{A\})^\om\times (\Si \cup\{A\})^\om$ such that $L \equiv_W R$.  
\nl It remains to consider the case of self dual Borel sets.  The alphabet $\Si$ being finite, a self dual Borel set $L$ is always Wadge equivalent to a Borel set 
in the form $\Si_1.L_1 \cup \Si_2.L_2$, where $(\Si_1, \Si_2)$ form a partition of $\Si$, and $L_1, L_2\subseteq \Sio$ are non self dual Borel sets such that 
$L_1 \equiv_W L_2^-$.  
Moreover $L_1$ and $L_2$ can be taken in the form $L_{(u_1)}=u_1.\Sio \cap L$ and      $L_{(u_2)}=u_2.\Sio \cap L$     for some $u_1, u_2 \in \Sis$, see
\cite{Duparc03}. 
So if  $L \subseteq \Sio$ is a self dual Borel set accepted by a  real time  B\"uchi $1$-counter automaton
 then $L \equiv_W \Si_1.L_1 \cup \Si_2.L_2$, where $(\Si_1, \Si_2)$ form a partition of $\Si$, and 
 $L_1, L_2\subseteq \Sio$ are non self dual Borel sets accepted by  real time  B\"uchi $1$-counter automata. 
We have already proved that there is  an infinitary rational relation $T_1 \subseteq \Gao \times \Gao$ 
 such that $T_1 \equiv_W L_1$ and an infinitary rational relation $T_2 \subseteq \Gao \times \Gao$ 
  such that $T_2 \equiv_W L_2$.  Thus 
$L \equiv_W  \Si_1.L_1 \cup \Si_2.L_2  \equiv_W \Ga_1.T_1\cup \Ga_2.T_2$, where $T_1$ and $T_2$ are subsets of 
$\Gao \times \Gao$ and $(\Ga_1, \Ga_2)$ form a partition of $\Ga \times \Ga$. Moreover 
 $\Ga_1.T_1 \cup \Ga_2.T_2$ is 
an infinitary rational relation. 
\ep 

\hs The Wadge hierarchy is a (great)  refinement of the Borel hierarchy and, 
 for each countable non null ordinal $\gamma$, ${\bf \Si}_\gamma^0$-complete sets (respectively, ${\bf \Pi}_\gamma^0$-complete sets) 
form a single equivalence class of $\equiv_W$, i.e. a single Wadge degree,  \cite{Wadge83,Kechris94}.  Thus we can state the following result which is a direct 
 consequence of  above Theorem \ref{thewad2} and of  \cite[Theorem 5.7]{Fin-mscs06}. 

\begin{Cor}\label{thebor2}    
\noi 
\begin{enumerate} 
\ite[(a)]  The Borel hierarchy of the class ${\bf RAT}_\om$ is equal to the Borel hierarchy of the class  $\Sigma^1_1$. 
\ite[(b)]   $\gamma_2^1$ is the supremum of the set of Borel ranks of infinitary rational relations.  
\ite[(c)]  For every non null  ordinal $\alpha < \om_1^{\mathrm{CK}}$, 
there exists some  
${\bf \Si}^0_\alpha$-complete and some ${\bf \Pi}^0_\alpha$-complete
infinitary rational relations.  
\end{enumerate}
\end{Cor}

\section{Concluding remarks}

\noi 
We have only considered above the Wadge hierarchy of {\bf  Borel sets}. If we assume the axiom of ${\bf \Si}_1^1$-determinacy, then Theorem 
\ref{thewad2} can be extended by considering the class of analytic sets instead of the class of Borel sets. In fact  in that case any set which is analytic but not Borel is 
${\bf \Si}_1^1$-complete, see \cite{Kechris94}. So there is only one more Wadge degree containing  ${\bf \Si}_1^1$-complete sets. 
We had already proved in \cite{Finkel03d} that 
there is a ${\bf \Si}_1^1$-complete set accepted by a $2$-tape  B\"uchi  automaton. 

\hs 
It is natural to ask for decidability results like : ``Is there an algorithm to determine the Wadge degree of a given infinitary rational relation accepted by a given 
$2$-tape B\"uchi automaton?". 
In the case of ($1$-tape)   automata the existence of such an algorithm has been proved by Wagner \cite{Wagner79}. 
\nl Unfortunately this is not possible in the case of  infinitary rational relations accepted by $2$-tape B\"uchi automata. 
We proved in \cite{Finkel03e} the following undecidability result : 

\begin{theorem}[\cite{Finkel03e}] Let  $\Si$ and $\Ga$ be finite alphabets having at least two letters,  and 
$\alpha$ be a countable ordinal $\geq 1$.  
Then for an effectively given infinitary rational relation $R \subseteq \Sio \times \Gao$ 
it is undecidable  to determine whether: 
\begin{enumerate}
\ite[(a)] 
$R$ is in the Borel class ${\bf \Si_{\alpha}^0}$.  
\ite[(b)] 
$R$ is in the Borel class ${\bf \Pi_{\alpha}^0}$.  
\ite[(c)] 
$R$ is a Borel subset  of $\Sio \times \Gao$. 
\ite[(d)] 
$R$ is a  ${\bf \Si^1_1}$-complete subset  of $\Sio \times \Gao$. 
\end{enumerate}

\end{theorem}

\noi This implies in particular that the Wadge hierarchy of  infinitary rational relations is  non effective.

\section*{Acknowledgements}

\hs Thanks to the anonymous referees for useful comments
on a preliminary version of this paper.


\begin{thebibliography}{BCPS03}

\bibitem[BCPS03]{BCPS03}
M.-P. B{\'e}al, O.~Carton, C.~Prieur, and J.~Sakarovitch.
\newblock Squaring transducers: an efficient procedure for deciding
  functionality and sequentiality.
\newblock {\em Theoretical Computer Science}, 292(1):45--63, 2003.

\bibitem[Ber79]{Berstel79}
J.~Berstel.
\newblock {\em Transductions and context free languages}.
\newblock Teubner Studienb\"ucher Informatik, 1979.

\bibitem[BT70]{BarzdinTrakhtenbrot}
Ya~M. Barzdin and B.A. Trakhtenbrot.
\newblock {\em Finite automata, behaviour and synthesis}.
\newblock Nauka, Moscow, 1970.
\newblock English translation, North Holland, Amsterdam, 1973.

\bibitem[B{\"u}c62]{Buchi62}
J.R. B{\"u}chi.
\newblock On a decision method in restricted second order arithmetic.
\newblock In Stanford~University Press, editor, {\em Proceedings of the 1960
  International Congress on Logic Methodology and Philosophy of Science}, pages
  1--11. Stanford University Press, 1962.

\bibitem[CG99]{ChoffrutGrigorieffG99}
C.~Choffrut and S.~Grigorieff.
\newblock Uniformization of rational relations.
\newblock In Juhani Karhum{\"a}ki, Hermann~A. Maurer, Gheorghe Paun, and
  Grzegorz Rozenberg, editors, {\em Jewels are Forever, Contributions on
  Theoretical Computer Science in Honor of Arto Salomaa}, pages 59--71.
  Springer, 1999.

\bibitem[CP97]{CartonPerrin97b}
O.~Carton and D.~Perrin.
\newblock Chains and superchains for {$\omega$}-rational sets, automata and
  semigroups.
\newblock {\em International Journal of Algebra and Computation},
  7(7):673--695, 1997.

\bibitem[CP99]{CartonPerrin99}
O.~Carton and D.~Perrin.
\newblock The {W}agner hierarchy of {$\omega$}-rational sets.
\newblock {\em International Journal of Algebra and Computation},
  9(5):597--620, 1999.

\bibitem[DFR01]{DFR}
J.~Duparc, O.~Finkel, and J.-P. Ressayre.
\newblock Computer science and the fine structure of {B}orel sets.
\newblock {\em Theoretical Computer Science}, 257(1--2):85--105, 2001.

\bibitem[DR06]{DR}
J.~Duparc and M.~Riss.
\newblock The missing link for $\omega$-rational sets, automata, and
  semigroups.
\newblock {\em  International Journal of Algebra and Computation},  16(1):161--186, 2006.


\bibitem[Dup01]{Duparc01}
J.~Duparc.
\newblock Wadge hierarchy and {V}eblen hierarchy: Part 1: Borel sets of finite
  rank.
\newblock {\em Journal of Symbolic Logic}, 66(1):56--86, 2001.

\bibitem[Dup03]{Duparc03}
J.~Duparc.
\newblock A hierarchy of deterministic context free $\omega$-languages.
\newblock {\em Theoretical Computer Science}, 290(3):1253--1300, 2003.

\bibitem[EH93]{eh}
J~Engelfriet and H.~J. Hoogeboom.
\newblock X-automata on $\om$-words.
\newblock {\em Theoretical Computer Science}, 110(1):1--51, 1993.

\bibitem[Fin01]{Fin01csl}
O.~Finkel.
\newblock An effective extension of the {W}agner hierarchy to blind counter
  automata.
\newblock In {\em Proceedings of Computer Science Logic, 15th International
  Workshop, CSL 2001}, volume 2142 of {\em Lecture Notes in Computer Science},
  pages 369--383. Springer, 2001.

\bibitem[Fin03a]{Finkel03f}
O.~Finkel.
\newblock On infinitary rational relations and {B}orel sets.
\newblock In {\em Discrete Mathematics and Theoretical Computer Science, 4th
  International Conference, DMTCS 2003, Dijon, France, July 7-12, 2003.
  Proceedings}, volume 2731 of {\em Lecture Notes in Computer Science}, pages
  155--167. Springer, 2003.

\bibitem[Fin03b]{Finkel03d}
O.~Finkel.
\newblock On the topological complexity of infinitary rational relations.
\newblock {\em RAIRO-Theoretical Informatics and Applications}, 37(2):105--113,
  2003.

\bibitem[Fin03c]{Finkel03e}
O.~Finkel.
\newblock Undecidability of topological and arithmetical properties of
  infinitary rational relations.
\newblock {\em RAIRO-Theoretical Informatics and Applications}, 37(2):115--126,
  2003.

\bibitem[Fin05]{cie05}
O.~Finkel.
\newblock Borel ranks and {W}adge degrees of context free $\omega$-languages.
\newblock In {\em Proceedings of New Computational Paradigms: First Conference
  on Computability in Europe, CiE 2005, Amsterdam, The Netherlands}, volume
  3526 of {\em Lecture Notes in Computer Science}, pages 129--138. Springer,
  2005.

\bibitem[Fin06a]{Fin-mscs06}
O.~Finkel.
\newblock Borel ranks and {W}adge degrees of omega context free languages.
\newblock {\em Mathematical Structures in Computer Science}, 16(5):813--840,
  2006.

\bibitem[Fin06b]{Fin06b}
O.~Finkel.
\newblock On the accepting power of two-tape {B}\"uchi automata.
\newblock In {\em Proceedings of the 23rd International Symposium on
  Theoretical Aspects of Computer Science, STACS 2006}, volume 3884 of {\em
  Lecture Notes in Computer Science}, pages 301--312. Springer, 2006.

\bibitem[Fin07]{Fin07-Mold}
O.~Finkel.
\newblock An example of ${\Pi}_3^0$-complete infinitary rational relation.
\newblock {\em Computer Science Journal of Moldova}, 15(1):3--21, 2007.

\bibitem[Gir81]{Gire-Phd}
F.~Gire.
\newblock {\em Relations rationnelles infinitaires}.
\newblock PhD thesis, Universit\'e Paris VII, 1981.

\bibitem[Gir83]{Gire83}
F.~Gire.
\newblock Une extension aux mots infinis de la notion de transduction
  rationelle.
\newblock In {\em Theoretical Computer Science, 6th GI-Conference, Dortmund,
  Germany, January 5-7, 1983, Proceedings}, volume 145 of {\em Lecture Notes in
  Computer Science}, pages 123--139. Springer, 1983.

\bibitem[GN84]{Gire-Nivat}
F.~Gire and M.~Nivat.
\newblock Relations rationnelles infinitaires.
\newblock {\em Calcolo}, pages 91--125, 1984.

\bibitem[Jec02]{Jech}
T.~Jech.
\newblock {\em Set theory, third edition}.
\newblock Springer, 2002.

\bibitem[Kec95]{Kechris94}
A.~S. Kechris.
\newblock {\em Classical descriptive set theory}.
\newblock Springer-Verlag, New York, 1995.

\bibitem[KMS89]{KMS89}
A.~S. Kechris, D.~Marker, and R.~L. Sami.
\newblock ${\Pi}_1^1$ {B}orel sets.
\newblock {\em Journal of Symbolic Logic}, 54(3):915--920, 1989.

\bibitem[Kur66]{Kuratowski66}
K.~Kuratowski.
\newblock {\em Topology}.
\newblock Academic Press, New York, 1966.

\bibitem[Lan69]{Landweber69}
L.H. Landweber.
\newblock Decision problems for $\omega$-automata.
\newblock {\em Mathematical Systems Theory}, 3(4):376--384, 1969.

\bibitem[LS77]{LindnerStaiger}
R.~Lindner and L.~Staiger.
\newblock {\em Algebraische codierungstheorie - Theorie der sequentiellen
  codierungen}.
\newblock Akademie-Verlag, Berlin, 1977.

\bibitem[LT94]{LescowThomas}
H.~Lescow and W.~Thomas.
\newblock Logical specifications of infinite computations.
\newblock In J.~W. de~Bakker, Willem~P. de~Roever, and Grzegorz Rozenberg,
  editors, {\em A Decade of Concurrency}, volume 803 of {\em Lecture Notes in
  Computer Science}, pages 583--621. Springer, 1994.

\bibitem[Mos80]{Moschovakis80}
Y.~N. Moschovakis.
\newblock {\em Descriptive set theory}.
\newblock North-Holland Publishing Co., Amsterdam, 1980.

\bibitem[Pin96]{Pin96}
J.-E. Pin.
\newblock Logic, semigroups and automata on words.
\newblock {\em Annals of Mathematics and Artificial Intelligence}, 16:343--384,
  1996.

\bibitem[PP04]{PerrinPin}
D.~Perrin and J.-E. Pin.
\newblock {\em Infinite words, automata, semigroups, logic and games}, volume
  141 of {\em Pure and Applied Mathematics}.
\newblock Elsevier, 2004.

\bibitem[Pri00]{PrieurPhd}
C.~Prieur.
\newblock {\em Fonctions Rationnelles de Mots Infinis et Continuit\'e}.
\newblock PhD thesis, Universit\'e Paris VII, 2000.

\bibitem[Sel95]{Selivanov95}
V.L. Selivanov.
\newblock Fine hierarchy of regular $\omega$-languages.
\newblock In {\em Proceedings of the International Joint Conference on the
  Theory and Practice of Software Development TAPSOFT-95, in Aarhus, Denmark},
  volume 915 of {\em Lecture Notes in Computer Science}, pages 277--287.
  Springer, 1995.

\bibitem[Sel98]{Selivanov98}
V.L. Selivanov.
\newblock Fine hierarchy of regular $\omega$-languages,.
\newblock {\em Theoretical Computer Science}, 191:37--59, 1998.

\bibitem[Sel03a]{Selivanov03a}
V.L. Selivanov.
\newblock Wadge degrees of $\omega$-languages of deterministic {T}uring
  machines.
\newblock In {\em Proceedings of the International Conference STACS 2003, 20th
  Annual Symposium on Theoretical Aspects of Computer Science, Berlin,
  Germany}, volume 2607 of {\em Lecture Notes in Computer Science}, pages
  97--108. Springer, 2003.

\bibitem[Sel03b]{Selivanov03b}
V.L. Selivanov.
\newblock Wadge degrees of $\omega$-languages of deterministic {T}uring
  machines.
\newblock {\em RAIRO-Theoretical Informatics and Applications}, 37(1):67--83,
  2003.

\bibitem[Sim92]{Simonnet92}
P.~Simonnet.
\newblock {\em Automates et th\'eorie descriptive}.
\newblock PhD thesis, Universit\'e Paris VII, 1992.

\bibitem[Sta86]{Staiger86a}
L.~Staiger.
\newblock Hierarchies of recursive $\omega$-languages.
\newblock {\em Elektronische Informationsverarbeitung und Kybernetik},
  22(5-6):219--241, 1986.

\bibitem[Sta97]{Staiger97}
L.~Staiger.
\newblock $\omega$-languages.
\newblock In {\em Handbook of formal languages, Vol.\ 3}, pages 339--387.
  Springer, Berlin, 1997.

\bibitem[SW78]{StaigerWagner78}
L.~Staiger and K.~Wagner.
\newblock Rekursive folgenmengen {I}.
\newblock {\em Z. Math Logik Grundlag. Math.}, 24:523--538, 1978.

\bibitem[Tho88]{Thomas88}
W.~Thomas.
\newblock Automata and quantifier hierarchies.
\newblock In {\em Formal Properties of Finite Automata and Applications, LITP
  Spring School on Theoretical Computer Science, Ramatuelle, France, May 25-29,
  1987, Proceedings}, volume 386 of {\em Lecture Notes in Computer Science},
  pages 104--119. Springer, 1988.

\bibitem[Tho90]{Thomas90}
W.~Thomas.
\newblock Automata on infinite objects.
\newblock In J.~van Leeuwen, editor, {\em Handbook of Theoretical Computer
  Science}, volume B, Formal models and semantics, pages 135--191. Elsevier,
  1990.

\bibitem[Wad83]{Wadge83}
W.~Wadge.
\newblock {\em Reducibility and determinateness in the Baire space}.
\newblock PhD thesis, University of California, Berkeley, 1983.

\bibitem[Wag79]{Wagner79}
K.~Wagner.
\newblock On $\omega$-regular sets.
\newblock {\em Information and Control}, 43(2):123--177, 1979.

\end{thebibliography}
\end{document}